\documentclass[useAMS,usenatbib]{mn2e}
\usepackage{graphicx}

\title[Dynamic spectrum of the Crab pulsar]
{Mechanism of generation of the emission bands\\ 
in the dynamic spectrum of the Crab pulsar}
\author[H. Ardavan et al.]{Houshang Ardavan,$^1$
Arzhang Ardavan,$^2$ John Singleton,$^{3}$ Mario Perez$^{4}$\\ 
$^1$Institute of Astronomy, University of Cambridge,
Madingley Road, Cambridge CB3 0HA, UK\\
$^2$Clarendon Laboratory, Department of Physics, University of Oxford,
Parks Road, Oxford OX1 3PU, UK\\
$^3$National High Magnetic Field Laboratory, MS-E536,
Los Alamos National Laboratory, Los Alamos, New Mexico 87545, USA\\
$^4$Space Sciences and Applications, ISR-1,
Los Alamos National Laboratory, Los Alamos, New Mexico 87545, USA}

\begin{document}

\date{18 April 2008}

\pagerange{\pageref{firstpage}--\pageref{lastpage}} \pubyear{2008}

\maketitle

\label{firstpage}

\begin{abstract}
We show that the proportionately spaced emission bands 
in the dynamic spectrum of the Crab pulsar 
(Hankins T.\ H.\ \& Eilek J.\ A., 2007, ApJ, 670, 693) 
fit the oscillations of the square of a Bessel function 
whose argument exceeds its order.  This function 
has already been encountered in the analysis of the 
emission from a polarization current with a 
superluminal distribution pattern: 
a current whose distribution pattern rotates 
(with an angular frequency $\omega$) and 
oscillates (with a frequency $\Omega>\omega$ 
differing from an integral multiple of $\omega$) 
at the same time (Ardavan H., Ardavan A. \& Singleton J., 
2003, J Opt Soc Am A, 20, 2137). 
Using the results of our earlier analysis, 
we find that the dependence on frequency of the spacing 
and width of the observed emission bands can be 
quantitatively accounted for by an appropriate choice of the 
value of the single free parameter $\Omega/\omega$. 
In addition, the value of this parameter, thus 
implied by Hankins \& Eilek's data, places the 
last peak in the amplitude of the oscillating 
Bessel function in question at a frequency 
($\sim\Omega^3/\omega^2$) that agrees 
with the position of the observed ultraviolet 
peak in the spectrum of the Crab pulsar.
We also show how the suppression of the emission 
bands by the interference of the contributions 
from differring polarizations can account 
for the differences in the time and 
frequency signatures of the interpulse and the 
main pulse in the Crab pulsar.  
Finally, we put the emission bands in the 
context of the observed continuum spectrum of the 
Crab pulsar by fitting this broadband spectrum 
(over 16 orders of magnitude of frequency) 
with that generated by an electric current with a 
superluminally rotating distribution pattern. 
\end{abstract}

\begin{keywords}
pulsars: individual (Crab Nebula pulsar)---radiation mechanisms: non-thermal.
\end{keywords}

\section{Introduction}
\label{sec:1}
Very soon after the discovery of pulsars,
it was realized that the 
very stable periodicity of the 
mean profiles of their
pulses could only result from
a source that {\it rotates}, and which
therefore posesses a rigidly rotating
radiation distribution~\citep{b25}.
In this paper, we show that this
source rotation is not only responsible
for the periodicity of the
pulses, but also determines the
detailed frequency dependence of the emitted radiation.
By inferring the values of two adjustable parameters 
from observational data (values that 
are consistent with those of plasma frequency and electron 
cyclotron frequency in a conventional pulsar magnetosphere), and 
by mildly restricting certain local properties 
of the source, we are able
to account {\it quantitatively} for the emission spectrum
of the Crab pulsar over 16 orders of
magnitude of frequency.

The rigid rotation of the overall distribution
pattern of the radiation from a pulsar is described by 
an electromagnetic field whose distribution depends on 
the azimuthal angle $\varphi$ only in the combination 
$(\varphi-\omega t)$,
where $\omega$ is the angular frequency of rotation of
the pulsar and $t$ is the time.
As we show in
Appendix A, Maxwell's equations demand that
the charge and current 
densities that give rise to this radiation field should 
have the same time dependence.
Therefore, the observed motion of the radiation pattern 
of pulsars can only arise 
from a source whose {\it distribution pattern} 
rotates rigidly, i.e.\ 
a source whose average 
density depends on $\varphi$ 
only in the combination $(\varphi-\omega t)$.
Furthermore, if a plasma distribution has a 
rigidly rotating pattern in the emission region, 
then it must have a rigidly rotating pattern 
everywhere; we show in Appendix B
that a solution 
of Maxwell's equations with the time dependence 
$\partial/\partial t=-\omega\partial/\partial\varphi$ 
applies either to the entire volume of the magnetospheric 
plasma distribution or to a region whose boundary is 
an expanding wave front that will eventually encompass 
the entire magnetosphere of the pulsar. 

Unless the plasma atmosphere surrounding the pulsar
is restricted to an unrealistically small volume,
it is therefore an inevitable consequence of the observational 
data that the macroscopic 
distribution of electric current in the magnetosphere 
has a rigidly rotating pattern 
whose linear speed exceeds the speed of 
light {\em in vacuo} $c$ for $r>c/\omega$, 
where $r$ is the radial distance from the axis of 
rotation.\footnote{A corollary of this 
implication is that the magnetospheric plasma 
of a pulsar cannot be fully charge-separated, 
as is commonly assumed in the works based on the 
Goldreich-Julian model \citep{b17}.}  
Although special relativity does not allow a charged
particle with a non-zero inertial mass to move faster than $c$, 
there is no such restriction on the speed of propagation 
of the variations in a macroscopic charge or current distribution.  
For instance, the distribution pattern of a
polarization current with a propagation speed that 
exceeds $c$ can be created by 
the coordinated motion of aggregates of particles that 
move slower than $c$~\citep{b1,b2,b3}.
Such a polarization-current density is on the same 
footing as the current density of free charges 
in the Amp\`{e}re-Maxwell equation, so that its 
propagating distribution pattern
radiates as would any other moving source 
of electromagnetic fields~\citep{b3};
indeed, such superluminal polarization currents 
have been demonstrated 
to be efficient emitters of
radiation in the laboratory~\citep{b4,b5,b16,b6}.

Once it is acknowledged that
the electric current emitting the observed pulses from pulsars
has a superluminally-rotating distribution pattern, 
results from the published literature 
on the electrodynamics of superluminal 
sources~\citep{b7,b8,b9,b10,b11,b12,b13} can be
applied to pulsars.
In this paper, we explain the recently-observed 
emission bands in the dynamic spectrum of the 
Crab pulsar~\citep{b14}
using the calculations of \citet{b8}.
Both the oscillations of the intensity
and the frequency dependence of the spacing and width 
of these bands are described by a Bessel function 
with a single parameter that is 
already constrained by other observational data on 
the spectrum of the Crab pulsar. 
This Bessel function is characteristic of the 
spectrum of the radiation by a superluminal 
polarization current whose distribution pattern 
rotates (with an angular frequency $\omega$) 
and oscillates (with a frequency $\Omega>\omega$ differing 
from an integral multiple of $\omega$) at the same time.
It differs from the Bessel function
encountered in the analysis of synchrotron radiation 
only in the relative magnitudes of its argument 
and its order: while the Bessel function describing 
synchrotron radiation has an argument
smaller than its order and so decays exponentially 
with increasing frequency, the Bessel 
function encountered in \citet{b8}, whose argument 
exceeds its order, is an oscillatory function of 
frequency with an amplitude that decays only algebraically.
It is this slower decay of the Bessel function 
in question that endows the emission from a 
superluminally-rotating source with a broad spectrum.
The physical mechanism underlying the broadband
nature of this emission is focusing in the time domain~\citep{b8,b12}:
contributions toward the intensity of the radiation
made over an extended period of emission 
time are received during a significantly 
shorter period of observation time.  

This paper is organized as follows. 
Section~\ref{newsec2} describes how the
most radiatively efficient parts of a pulsar magnetosphere are thin 
filaments within the superluminally rotating part of 
its current distribution pattern~\citep{b10}.
A knowledge of the morphology of these filaments~\citep{b10}, and the
ability of a source that travels faster than its own
waves to make multiple contributions to the
signal received by an observer at a given instant~\citep{b9}, are necessary to
understand many of the
traits of pulsar observations described later in the paper. 
Section~\ref{newsec3} summarizes our 
earlier analysis \citep{b8} on the frequency 
spectrum of the radiation from a rotating superluminal source.  
Using these results, we derive various features of the 
emission bands in Section~\ref{newsec4} and compare
them to the frequency bands seen in the interpulses
of the Crab pulsar [Figs.~6--8, \citet{b14}];
using a single input parameter,
related to the pulsar's plasma density, we
reproduce the observational bands and
predict the final ultra-violet emission peak of the Crab pulsar.
Section~\ref{newsec5} discusses the suppression 
of the bands by interference; this is relevant to
the much less frequency-banded microbursts and nanoshots
of the Crab's main pulses [Figs.~2 and 3, \citet{b14}].
Section~\ref{newsec6} describes the continuum spectrum 
of the Crab pulsar; by introducing a further input
parameter related to the plasma dynamics we are
able to account {\it quantitatively} for the
whole emission spectrum over 16 orders of magnitude
of frequency.
Section~\ref{newsec7} gives a short discussion and summary.
The mathematical details of our arguments are presented in 
Appendices A, B and C.  

\section{The emitting region of a 
rotating superluminal source: multivalued nature of the retarded time }
\label{newsec2}
\begin{figure}
\includegraphics[width=8cm]{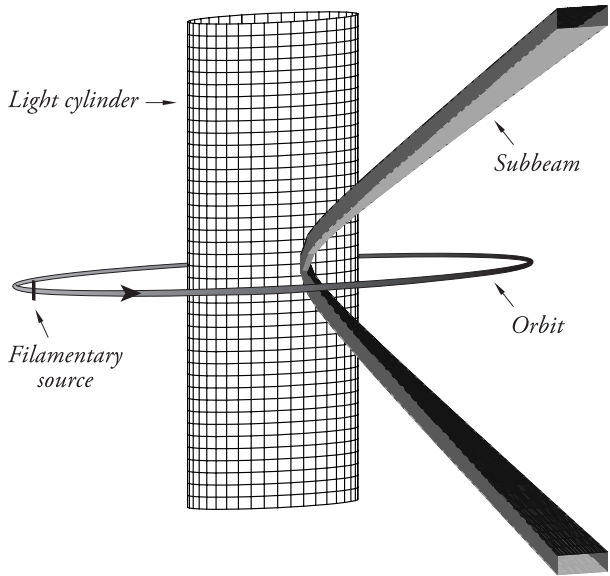}
\caption{Schematic illustration of the light 
cylinder $r=c/\omega$, the filamentary part of the 
distribution pattern of the source that approaches the 
observation point with the speed of light and 
zero acceleration at the retarded time, 
the orbit of this filamentary source, and the subbeam 
formed by the bundle of cusps that emanate 
from the constituent volume elements of this filament
[after \citet{b10}].
The subbeam is diffractionless in the polar 
direction.  The figure represents a snapshot corresponding 
to a fixed value of the observation time $t_P$.}
\label{fig1}   
\end{figure}

At large distances from the source,
the radiation field of a superluminally rotating 
extended source at an observation point $P$ 
is dominated by the emissions of its volume 
elements that approach $P$ along the radiation direction 
with the speed of light and zero acceleration 
at the retarded time~\citep{b9,b10}.
These elements constitute
a filamentary part of the source 
whose radial and azimuthal widths become narrower 
(as $\delta r\sim{R_P}^{-2}$ and $\delta\varphi\sim{R_P}^{-3}$, 
respectively), the larger the distance $R_P$ of 
the observer from the source, and whose length is 
of the order of the length scale $l_z$ of the source 
parallel to the axis of rotation~\citep{b10}.  
(Here, $r$, $\varphi$, and $z$ are the cylindrical 
polar coordinates of the source points.) 
For an observation point $P$ with spherical polar
coordinates ($R_P,\varphi_P, \theta_P$) that is 
located in the far zone, this contributing part of 
the source lies at ${\hat r}=\csc\theta_P, \varphi=\varphi_P+3\pi/2$,
and is essentially a straight line parallel 
to the rotation axis, as shown in Fig.~\ref{fig1}~\citep{b10}.
(The dimensionless coordinate ${\hat r}$ 
stands for $r\omega/c$, where $\omega$, 
the angular frequency of rotation of the distribution
pattern of the source, is the same frequency as that
with which the pulsar rotates.)

\begin{figure}
\includegraphics[width=8cm]{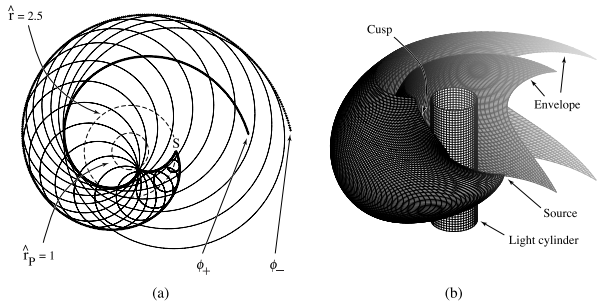}
\caption{(a) Cross section of the \v Cerenkov-like envelope 
(bold curves) of the spherical Huygens wave fronts 
(fine circles) emitted by a volume element 
$S$ within the distribution pattern of an 
extended, rotating superluminal source of 
angular frequency $\omega$
[after \citet{b9}]. The point $S$ 
is on a circle of radius $r=2.5\,c/\omega$, 
or ${\hat r}\equiv r\omega/c=2.5$, 
i.e.\ has an instantaneous linear velocity 
$r\omega=2.5\,c$. The cross section is 
with the plane of $S$'s orbit; dashed 
circles designate the light cylinder 
$r_P=c/\omega$ (${\hat r}_P=1$) and the 
orbit of $S$. (b) Three-dimensional view of 
the light cylinder, the envelope of wave 
fronts emanating from $S$, and the cusp 
along which the two sheets of this envelope 
meet tangentially. The cusp is tangent to 
the light cylinder in the plane of the orbit and 
spirals outward into the far zone.}
\label{fig2}   
\end{figure}

Once a source travels faster than its emitted waves, 
it can make more than one retarded contribution to 
the field observed at any given instant~\citep{b8,b9,b12}.
This multivaluedness of the retarded time means that 
the wave fronts emitted by each of the contributing 
elements of the source possess an envelope, which in 
this case consists of a two-sheeted, tube-like surface 
whose sheets meet tangentially along a spiraling 
cusp curve; see Fig.~\ref{fig2}~\citep{b9}. 
For moderate superluminal speeds, 
the field inside the envelope receives contributions 
from three distinct values of the retarded time, 
while the field outside the envelope is influenced 
only by a single instant of emission time~\citep{b11}.  
Coherent superposition of the emitted waves on the 
envelope (where two of the contributing retarded times coalesce) 
and on its cusp (where all three of the contributing 
retarded times coalesce) results in not only a 
spatial but also a temporal focusing of the waves: 
the contributions from emission over an extended 
period of retarded time reach an observer who is located 
on the cusp during a significantly shorter 
period of observation time~\citep{b12}.

The field of each contributing volume element of the source 
is strongest, therefore, on the cusp of the envelope 
of wave fronts that it emits
[see \citet{b10} and references therein].  
The bundle of cusps 
generated by the collection of the contributing 
source elements (i.e.\ by the filamentary part 
of the source that approaches the observer with the 
speed of light and zero acceleration) constitutes a 
radiation subbeam whose widths in the polar and 
azimuthal directions are of the order of 
$\delta\theta_P\sim{R_P}^{-1}$ and $\delta\varphi_P\sim{R_P}^{-3}$, 
respectively~\citep{b10}.
The overall radiation beam generated by 
the source consists of a (necessarily 
incoherent\footnote{The superposition of the subbeams is 
necessarily incoherent because the subbeams that are 
detected at two neighbouring points within the 
overall beam arise from two distinct 
filamentary parts of the source with essentially 
no common elements.  The incoherence of this 
superposition would ensure that, though the 
field amplitude within a subbeam, which 
narrows with distance, decays nonspherically, the 
field amplitude associated with the overall radiation 
beam, which occupies a constant solid angle, does not.}) 
superposition of such subbeams. This beam's azimuthal 
width is the same as the azimuthal extent of the source, 
and its polar width, 
$\arccos(1/{\hat r}_<)\le\vert\theta_P-\pi/2\vert\le\arccos(1/{\hat r}_>)$, 
is determined by the radial extent 
$1<{\hat r}_<\le{\hat r}\le{\hat r}_>$ of the 
superluminal part of the source~\citep{b10}.
This will be important in Secs.~\ref{newsec4}
and \ref{newsec5}; the $\varphi_P$ 
dependence of the radiation intensity within the 
overall beam (i.e.\ what is observed as the main pulse, 
interpulse, and other components of the mean pulse) 
thus reflects the distribution of the source density 
around the cylindrical surface 
${\hat r}=\csc\theta_P$ from which 
the main contributions to the field arise.

Since the cusps represent the loci of points at which 
the emitted spherical waves interfere constructively 
[i.e.\ represent wave packets that are constantly 
dispersed and reconstructed out of other waves~\citep{b7}], 
the subbeams generated by a superluminal source need 
not be subject to diffraction as are conventional radiation beams.
Nevertheless, they have a decreasing angular width 
(i.e.\ are nondiffracting) only in the polar direction~\citep{b10}. 
Their azimuthal width $\delta\varphi_P$ decreases as 
${R_P}^{-3}$ with distance because they 
receive contributions from an azimuthal extent 
$\delta\varphi$ of the source that 
likewise shrinks as ${R_P}^{-3}$.
They would have had a constant azimuthal width
had the azimuthal extent of the contributing 
part of the source been independent of $R_P$.
On the other hand, the solid angle occupied 
by the cusps has a thickness $\delta z_P$ in the 
direction parallel to the rotation axis that remains 
of the order of the height $l_z$ of the source distribution 
at all distances (see Fig.~1).
Consequently, the polar width $\delta\theta_P$ of the 
particular subbeam that goes through the observation
point decreases as ${R_P}^{-1}$, instead of being 
independent of $R_P$~\citep{b10}.

Because it has a constant linear width parallel
to the rotation axis, an individual subbeam subtends 
an area of the order of $R_P$, rather than ${R_P}^2$.
In order that the energy flux remain the same across 
all cross sections of the subbeam, therefore, 
it is essential that the Poynting vector associated 
with this radiation correspondingly decay more slowly 
than that of a conventional, spherically decaying 
beam: as ${R_P}^{-1}$, rather than ${R_P}^{-2}$, 
within the bundle of cusps that emanate from the 
constituent volume elements of the source and extend 
into the far zone~\citep{b9}. This result, which also 
follows from the superposition of the 
Li\'enard-Wiechert fields of the constituent 
volume elements of a rotating superluminal 
source~\citep{b10}, has been 
demonstrated experimentally~\citep{b5,b16}.

The fact that the observationally-inferred dimensions of 
the plasma structures responsible for the 
emission from pulsars are less than $1$ metre 
in size \citep{b21} reflects, in the present 
context, the narrowing (as ${R_P}^{-2}$ and ${R_P}^{-3}$, 
respectively) of the radial and azimuthal 
dimensions of the filamentary part of the 
source that approaches the observer with the speed 
of light and zero acceleration at the retarded time.
Not only do the nondiffracting subbeams that 
emanate from such filaments account for the nanostructure,
and so the brightness temperature, of the giant 
pulses, but the nonspherical decay of the intensity 
of such subbeams (as ${R_P}^{-1}$ instead 
of ${R_P}^{-2}$) explains why their energy densities 
at their source appear to exceed the energy densities 
of both the plasma and the magnetic field at 
the suface of a neutron star when estimated 
on the basis of the inverse square law \citep{b22}. 

\section[]{Frequency spectrum of the radiation 
from a rotating superluminal source}
\label{newsec3}
In this section we summarize
the results of \citet{b8} relevant to pulsars;
the original intent of that paper
was to calculate the spectrum of the 
radiation from a generic superluminal source 
that has been implemented in the 
laboratory~\citep{b5,b16}. 
This source comprises a polarization-current 
density ${\bf j}=\partial{\bf P}/\partial t$
for which
\begin{eqnarray}
P_{r,\varphi,z}(r,\varphi,z,t)
&=&s_{r,\varphi,z}(r,z)\cos(m\hat{\varphi})\cos(\Omega t),\nonumber\\
&&\qquad\qquad\qquad\qquad-\pi<\hat{\varphi}\le\pi,
\label{eq:1}
\end{eqnarray}
with
\begin{equation}
\hat{\varphi}\equiv\varphi-\omega t,
\label{eq:2}
\end{equation}
where $P_{r,\varphi,z}$ are the components 
of the polarization ${\bf{\it P}}$
in a cylindrical coordinate system ($r,\varphi,z$) based on the axis 
of rotation,
${\bf s}(r,z)$ is an arbitrary vector
that vanishes outside a finite region of the $(r,z)$ space,
and $m$ is a positive integer.
For a fixed value of $t$,
the azimuthal dependence of the polarization (\ref{eq:1})
along each circle of radius $r$ within the source
is the same as that of a sinusoidal wave 
train with the wavelength $2\pi r/m$
whose $m$ cycles fit around the 
circumference of the circle smoothly.
As time elapses,
this wave train both propagates around each 
circle with the velocity $r\omega$
and oscillates in its amplitude with the frequency $\Omega$.
This is a generic source: 
one can construct any distribution with 
a uniformly rotating pattern,
$P_{r,\varphi,z}(r,\hat{\varphi},z)$ or $j_{r,\varphi,z}(r,\hat{\varphi},z)$,
by the superposition over $m$
of terms of the form $s_{r,\varphi,z}(r,z,m)\cos(m\hat{\varphi})$.  
In the following discussion, we assume
that the modulation frequency $\Omega$ is positive and different 
from an exact integral multiple of the 
rotation frequency $\omega$ [for the significance 
of this incommensurablity requirement, see \citet{b8}].  

Although formulated in terms of a polarization current 
(for which there is manifestly no restriction on the 
propagation speed of the variations in the current density),
the results that follow hold true for any current 
distribution whose density depends on 
$\varphi$ as $\varphi-\omega t$ in $r>c/\omega$.  The electromagnetic fields
\begin{equation}
{\bf E}=-{\bf\nabla}_P A^0 - c^{-1} \partial{\bf A}/\partial t_P,
\quad {\bf B}={\bf\nabla}_P{\bf\times A},
\label{eq:3}
\end{equation}
that arise from such a source are given,
in the absence of boundaries,
by the following classical expression for the retarded four-potential:
\begin{eqnarray}
A^\mu({\bf x}_P,t_P)&=&c^{-1}\int{\rm d}^3 x{\rm d} t\, j^\mu({\bf x},t)\delta(t_P-t-R/c)/R\nonumber\\
&&\qquad\qquad\qquad\qquad\qquad \mu=0,\cdots,3.
\label{eq:4}
\end{eqnarray}
Here,
$({\bf x}_P,t_P)=(r_P,\varphi_P,z_P,t_P)$ and $({\bf x},t)=(r,\varphi,z,t)$
are the space-time coordinates of the observation 
point and the source points, respectively,
$R$ stands for the magnitude of ${\bf R}\equiv{\bf x}_P-{\bf x}$,
and $\mu=1,2,3$ designate the spatial components,
${\bf A}$ and ${\bf j}$,
of $A^\mu$ and $j^\mu$ in a 
Cartesian coordinate system \citep{b18}.

In \citet{b8},
we first calculated the Li\'enard-Wiechert field
that arises from a circularly moving point source
(representing a volume element of an extended source)
with a superluminal speed $r\omega>c$,
i.e.\ considered a generalization
of the synchrotron radiation to the superluminal regime.
We then evaluated the integral representing the retarded field
of the extended source (\ref{eq:1})
by superposing the fields
generated by the constituent volume elements of this source,
i.e.\ by using the generalization of the synchrotron field
as the Green's function for the problem
[see equation (19) of \citet{b8}].
For a source that travels faster than $c$,
this Green's function has extended singularities
arising from the constructive interference
of the emitted waves on the 
envelope of wave fronts and its cusp (see Fig.~\ref{fig2}).

Correspondingly, the fields ${\bf E}$ and ${\bf B}$ 
are given, in the frequency domain, by 
radiation integrals with rapidly oscillating 
integrands whose phases are stationary on the 
loci of the coherently contributing source 
elements: those source elements that approach 
the observer, along the radiation direction, 
with the speed of light and zero acceleration at the 
retarded time. For a radiation angular frequency 
$n\omega$ that appreciably exceeds the 
angular velocity $\omega$,\footnote{When discussing
pulsars in the following sections, we shall be
considering radiation frequencies in the
GHz-ultaviolet range, whereas the rotation frequency
of the Crab pulsar is $\omega /2\pi \simeq 30$~Hz.
The condition $n\omega \gg \omega$ will 
therefore always hold.}
these integrals can be evaluated 
by the method of stationary phase to obtain the 
following expression for the electric field of 
the emitted radiation outside the plane of the source's orbit:
\begin{equation}
{\bf E}=\Re\left\{{\tilde{\bf E}}_0+2\sum_{n=1}^\infty{\tilde{\bf E}}_n\exp(-{\rm i}n{\hat\varphi}_P)\right\},
\label{eq:5}
\end{equation}
in which
\begin{eqnarray}
{\tilde{\bf E}}_n&\simeq&{\textstyle\frac{1}{2}}{{\hat r}_P}^{-1}\exp\{-{\rm i}[n({\hat R}_P+{\textstyle\frac{3}{2}}\pi)-(\Omega/\omega)(\varphi_P+{\textstyle\frac{3}{2}}\pi)]\}\nonumber\\
&&\times Q_{\hat\varphi}{\bar Q}_r{\bar{\bf Q}}_z+\{m\to -m,\Omega\to-\Omega\},
\label{eq:6}
\end{eqnarray}
\begin{equation}
Q_{\hat\varphi}=(-1)^{n+m}\sin\left(\frac{\pi\Omega}{\omega}\right)\left(\frac{{\mu_+}^2}{n-\mu_+}+\frac{{\mu_-}^2}{n-\mu_-}\right),
\label{eq:7}
\end{equation}
\begin{equation}
{\bar Q}_r\simeq{\hat r}_>-{\hat r}_<,\qquad n\ll\pi{\hat R}_P/({\hat r}_>-{\hat r}_<)^2,
\label{eq:8}
\end{equation}
or
\begin{equation}
{\bar Q}_r\simeq(2\pi{\hat R}_P/n)^{1/2}\exp(-{\rm i}\pi/4),\qquad n\gg\pi{\hat R}_P/({\hat r}_>-{\hat r}_<)^2,
\label{eq:8a}
\end{equation}
and
\begin{eqnarray}
{\bar{\bf Q}}_z&=&\big[{\bar s}_r{\bf J}_{n-\Omega/\omega}(n)+{\rm i}{\bar s}_\varphi{{\bf J}^\prime}_{n-\Omega/\omega}(n)\big]{\hat{\bf e}}_\parallel\nonumber \\
&&+\big[({\bar s}_\varphi\cos\theta_P
-{\bar s}_z\sin\theta_P){\bf J}_{n-\Omega/\omega}(n)\nonumber\\
&&\qquad-{\rm i}{\bar s}_r\cos\theta_P{{\bf J}^\prime}_{n-\Omega/\omega}(n)\big]{\hat{\bf e}}_\perp,
\label{eq:9}
\end{eqnarray}
with
\begin{equation}
{\bar s}_{r,\varphi,z}\equiv\int_{-\infty}^\infty {\rm d}{\hat z}\,\exp({\rm i}n{\hat z}\cos\theta_P)s_{r,\varphi,z}\big\vert_{{\hat r}=\csc\theta_P}
\label{eq:10}
\end{equation}
[see equations (15), (23), (46) and (66)--(68) of \citet{b8}].  
Here, $({\hat r},{\hat z};{\hat r}_P,{\hat z}_P)$ 
stand for $(r\omega/c, z\omega/c;r_P\omega/c, z_P\omega/c)$, 
$(R_P,\theta_P,\varphi_P)$ are the spherical 
polar coordintes of the observation point 
$P$, $\mu_\pm\equiv(\Omega/\omega)\pm m$, 
${\hat r}_<<1$ and ${\hat r}_>>1$ are the upper 
and lower limits of the radial interval in which 
the source densities $s_{r,\varphi,z}$ are non-zero, 
${\bf J}$ and ${\bf J}^\prime$ are the 
Anger function and the derivative of the 
Anger function with respect to its argument, 
respectively, and ${\hat{\bf e}}_\parallel\equiv{\hat{\bf e}}_z\times{\hat{\bf n}}/|{\hat{\bf e}}_z\times{\hat{\bf n}}|$
(which is parallel to the plane of rotation)
and ${\hat{\bf e}}_\perp\equiv{\hat{\bf n}}{\bf\times}{\hat{\bf e}}_\parallel$
comprise a pair of unit vectors normal to the radiation direction 
${\hat{\bf n}}$ (${\hat{\bf e}}_z$ is the 
base vector associated with the coordinate $z$).
The symbol $\{m\to -m,\Omega\to-\Omega\}$ designates 
a term exactly like the one preceding 
it but in which $m$ and $\Omega$ are everywhere 
replaced by $-m$ and $-\Omega$, respectively. 

The radiated power per harmonic per unit 
solid angle is therefore given by
\begin{eqnarray}
{\rm d}P_n/{\rm d}\Omega_P&=&(2\pi)^{-1}c{R_P}^2|{\tilde{\bf E}}_n|^2\nonumber\\
&\simeq&(8\pi)^{-1}(c^3/\omega^2)\csc^2\theta_P\vert{\bar Q}_r\vert^2{Q_{\hat\varphi}}^2\vert {\bar{\bf Q}}_z\vert^2,
\label{eq:11}
\end{eqnarray}
where ${\rm d}\Omega_P$ denotes the element 
$\sin\theta_P{\rm d}\theta_P{\rm d}\varphi_P$ of 
solid angle in the space of observation points.
The contribution of the term 
$\{m\to -m,\Omega\to-\Omega\}$ in equation 
(\ref{eq:6}) has been ignored here because 
the Anger functions for $-\Omega<0$ turn out 
to be exponentially smaller than those 
for positive $\Omega$ (see Appendix C). 

There is no difference between an Anger function, 
${\bf J}_\nu(\chi)$, and a Bessel function of 
the first kind, $J_\nu(\chi)$, when $\nu$ is an integer.
Even for a non-integral value of $\nu$, 
the difference between these two functions vanishes, 
as $\chi^{-1}$, if $\chi\gg1$.
In the regime $n\gg1$, where the radiation 
frequency $n\omega$ appreciably exceeds 
the rotation frequency, therefore, 
the Anger functions ${\bf J}_{n-\Omega/\omega}(n)$ 
and ${{\bf J}^\prime}_{n-\Omega/\omega}(n)$ 
in Equation (\ref{eq:9}) can be replaced by the 
Bessel functions $J_{n-\Omega/\omega}(n)$ and
${J^\prime}_{n-\Omega/\omega}(n)$, respectively.
If the radiation frequency $n\omega$ appreciably 
exceeds also the modulation frequency 
$\Omega$, then these Bessel functions can in 
turn be approximated by the following Airy functions:
\begin{equation}
J_{n-\frac{\Omega}{\omega}}(n)\simeq\left(\frac{2}{n}\right)^{1/3}{\rm Ai}\left[-\left(\frac{2}{n}\right)^{1/3}\frac{\Omega}{\omega}\right],
\label{eq:12}
\end{equation}
\begin{equation}
{J^\prime}_{n-\frac{\Omega}{\omega}}(n)\simeq-\left(\frac{2}{n}\right)^{2/3}{\rm Ai}^\prime\left[-\left(\frac{2}{n}\right)^{1/3}\frac{\Omega}{\omega}\right],
\label{eq:13}
\end{equation}
where ${\rm Ai}^\prime$ stands for the derivative of 
the Airy function with respect to its argument (see Appendix C).
 
Hence, the radiated power is given by
\begin{eqnarray}
\frac{{\rm d}P_n}{{\rm d}\Omega_P}&\simeq&\frac{c^3}{8\pi^2\omega^2}\csc^2\theta_P\vert{\bar Q}_r\vert^2{Q_{\hat\varphi}}^2\nonumber\\
&&\times\left(\frac{2}{n}\right)^{2/3}\Bigg\vert\left[{\bar s}_r{\hat{\bf e}}_\parallel+({\bar s}_\varphi\cos\theta_P-{\bar s}_z\sin\theta_P){\hat{\bf e}}_\perp\right]\nonumber\\
&&\times{\rm Ai}\left[-\left(\frac{2}{n}\right)^{1/3}\frac{\Omega}{\omega}\right]-{\rm i}({\bar s}_\varphi{\hat{\bf e}}_\parallel+{\bar s}_r\cos\theta_P{\hat{\bf e}}_\perp)\nonumber\\
&&\times\left(\frac{2}{n}\right)^{1/3}{\rm Ai}^\prime\left[-\left(\frac{2}{n}\right)^{1/3}\frac{\Omega}{\omega}\right]\Bigg\vert^2
\label{eq:14}
\end{eqnarray}
in the regime where the radiation frequency ($n\omega$) 
apprecialy exceeds both the rotation 
($\omega$) and modulation ($\Omega$) frequencies 
of the source distribution (\ref{eq:1}).  

Asymptotic values of the amplitudes of 
${\rm Ai}[-(2/n)^{1/3}(\Omega/\omega)]$ and 
${\rm Ai}^\prime[-(2/n)^{1/3}(\Omega/\omega)]$ 
are respectively given by $(2/n)^{-1/12}(\Omega/\omega)^{-1/4}$ 
and $(2/n)^{1/12}(\Omega/\omega)^{1/4}$, 
when $n\ll(\Omega/\omega)^3$, and by 
the constants ${\rm Ai}(0)$ and ${\rm Ai}^\prime(0)$ 
when $n\gg(\Omega/\omega)^3$ \citep{b19}. 
Moreover, the quantity $Q_{\hat\varphi}$, which 
appears in equation (\ref{eq:14}), 
is independent of $n$ ($\sim2\Omega/\omega$) 
if $n\ll\mu_+$, decays as $n^{-1}$ if $n\gg\mu_+$, 
and is of the order of $n$ if 
$n\simeq m\gg\Omega/\omega$ [see equation(\ref{eq:7})].

Depending on the relative magnitudes of 
$n$, $m$ and $\Omega/\omega$, therefore, 
the amplitude of the radiated power ${\rm d}P_n/{\rm d}\Omega_P$ 
decays with the harmonic number $n$ as 
$n^{-\sigma}|{\bar s}_{r,\varphi,z}(n)|^2$, 
where the exponent $\sigma$ can assume one of 
the values $1/2$, $2/3$, $5/2$ or $8/3$, and 
${\bar s}_{r,\varphi,z}(n)$ denotes the frequency 
dependence of the Fourier components of the 
source densities $s_{r,\varphi,z}$.
The way in which ${\bar s}_{r,\varphi,z}(n)$ 
decay with $n$, for large $n$, is determined 
by the distribution of the source density ${\bf s}(r,z)$
with respect to $z$ [see equation (\ref{eq:10})], 
and so is model-dependent.\footnote{Note that the electric 
susceptibility of the magnetospheric 
plasma (contained in the factor ${\bf s}$) 
does not influence these results, because 
it depends on the source frequencies 
$\mu_\pm\omega$, not on the radiation frequency 
$n\omega$ [see \citet{b13}]}  
However, the spacings between the emission bands 
predicted by equation (\ref{eq:14}) turn out to be 
essentially independent of the rate of decay of the 
amplitudes of these bands (see below).

Having assembled the relevant equations,
we now compare their predictions with
established observational data on the Crab pulsar.

\begin{figure}
\includegraphics[width=8cm]{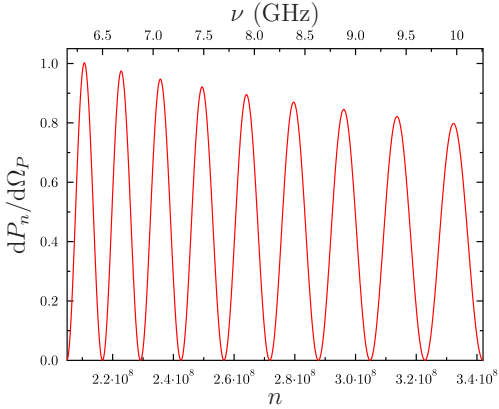}
\caption{The power ${\rm d}P_n/{\rm d}\Omega_P$ radiated per 
harmonic per unit solid angle (in arbitrary units), 
into the signal linearly polarized parallel to the 
plane of rotation, versus the harmonic number 
$n$ (lower axis) for $\Omega/\omega=1.9\times10^4$, 
$a_\parallel=1.48\times10^{-5}$ and $b_\parallel=1.03\times10^4$.
The upper axis gives
observation frequency $n\omega /2 \pi$ for the
case of the Crab pulsar [$\omega/(2\pi)\simeq 30$~Hz].
The spacing, the width, and the number of oscillations
in this figure are the same as those of 
the bands seen in the
interpulse data from the Crab pulsar over 
the frequency interval $8<\nu<10.5$~GHz
[see Figs.~6--8; \citet{b14}].}
\label{fig3}
\end{figure}
 
\section{Predicted frequency bands: comparison with
the spectra of the Crab interpulses} 
\label{newsec4}
One of the most remarkable features of the observational
data from the Crab pulsar is the presence
of well-defined
frequency bands within the midpulses~\citep{b14}.
We now show that these are a natural consequence of
a rotating superluminal source and use their
spacing to predict the high-frequency
emission of the Crab pulsar.

When the absolute value of their argument 
appreciably exceeds unity, the Airy functions 
in equation (\ref{eq:14}) are rapidly oscillating 
functions of the harmonic number $n$ with peaks 
that spread further apart with increasing 
$n$ (Fig.~\ref{fig3}). 
The rate of increase of the spacing between the peaks 
in question is determined solely by the value of 
$\Omega/\omega$. 
We now derive an explicit expression for the spacing
$\Delta n$ of the maxima of the radiated
power ${\rm d}P_n/{\rm d}\Omega_P$ as a 
function of $n$ and compare the result with 
the proportionately spaced emission 
bands observed by \citet{b14}.

Within the range $\Omega/\omega\ll n\ll(\Omega/\omega)^3$ 
of harmonic numbers, where they are oscillatory, 
the Airy functions in equation (\ref{eq:14}) 
can be further approximated by
\begin{equation}
{\rm Ai}[-(2/n)^{1/3}(\Omega/\omega)]\simeq\pi^{-1/2}(n/2)^{1/12}(\Omega/\omega)^{-1/4}\cos\psi  
\label{eq:14a}
\end{equation}
and
\begin{equation}
{\rm Ai}^\prime[-(2/n)^{1/3}(\Omega/\omega)]\simeq\pi^{-1/2}(n/2)^{-1/12}(\Omega/\omega)^{1/4}\sin\psi 
\label{eq:14b}
\end{equation}
where
\begin{equation}
\psi\equiv\textstyle{2\over3}(2/n)^{1/2}(\Omega/\omega)^{3/2}-\pi/4 
\label{eq:14c}
\end{equation}
(see Appendix C).  In this regime, therefore, 
equation (\ref{eq:14}) reduces to
\begin{equation}
\frac{{\rm d}P_n}{{\rm d}\Omega_P}\simeq\frac{c^3\csc^2\theta_P}{8\pi^2\omega^2}\vert{\bar Q}_r\vert^2{Q_{\hat\varphi}}^2
\left(\frac{2\omega}{n\Omega}\right)^{1/2}\big\vert{\cal P}_\parallel{\hat{\bf e}}_\parallel+{\cal P}_\perp{\hat{\bf e}}_\perp\big\vert^2,
\label{eq:15}
\end{equation}
in which
\begin{equation}
{\cal P}_\parallel\equiv{\bar s}_r\cos\psi-{\rm i}\left(\frac{2\Omega}{n\omega}\right)^{1/2}{\bar s}_\varphi\sin\psi
\label{eq:15a}
\end{equation}
and
\begin{equation}
{\cal P}_\perp\equiv({\bar s}_\varphi\cos\theta_P-{\bar s}_z\sin\theta_P)\cos\psi-{\rm i}\frac{2\Omega}{n\omega}{\bar s}_r\cos\theta_P\sin\psi
\label{eq:15b}
\end{equation}
represent the contributions towards the radiated 
field ${\tilde{\bf E}}_n$ from differing polarizations.

Let us first consider the contribution 
towards the radiated power from the component 
of the field parallel to the plane of rotation.  
It follows from equation (\ref{eq:15a}) that
\begin{eqnarray}
\vert{\cal P}_\parallel\vert^2
&=&\sec(2\gamma_\parallel)\big\{-\vert{\bar s}_r\vert^2\sin^2\gamma_\parallel+\frac{2\Omega}{n\omega}\vert{\bar s}_\varphi\vert^2\cos^2\gamma_\parallel\nonumber\\
&&+[\vert{\bar s}_r\vert^2-\frac{2\Omega}{n\omega}\vert{\bar s}_\varphi\vert^2]\cos^2(\psi-\gamma_\parallel)\big\},
\label{eq:15c}
\end{eqnarray}
where
\begin{equation}
\gamma_\parallel=\textstyle{1\over2}\arctan\frac{[2\Omega/(n\omega)]^{1/2}\Im\{{\bar s}_r{{\bar s}_\varphi}^*\}}{\vert{\bar s}_r\vert^2-[2\Omega/(n\omega)]\vert{\bar s}_\varphi\vert^2},
\label{eq:15d}
\end{equation}
with $\Im$ and $*$ denoting the imaginary part 
and a complex conjugate, respectively.
Two successive oscillations of $\vert{\cal P}_\parallel\vert^2$ 
in the neighbourhood of any given $n$ arise, 
according to equation (\ref{eq:15c}), 
from a change in the value of the argument of 
$\cos^2(\psi-\gamma_\parallel)$ by $\pi$.
Therefore, the spacing between two successive peaks 
of this function is locally given by
\begin{equation}
\Delta n=\pi\big\vert\frac{\partial}{\partial n}(\psi-\gamma_\parallel)\big\vert^{-1},
\label{eq:15e}
\end{equation}
an equation that yields $\Delta n$ versus $n$ for 
all $n$.\footnote{At any point
such equations may be readily translated into the frequency units
$\Delta \nu$ versus $\nu$ employed in Fig.~10 of
\citet{b14} by multiplying both $\Delta n$ and
$n$ by $\omega/2\pi\simeq 30$~Hz,
the rotation frequency relevant for the
Crab pulsar.}

If the source densities ${\bar s}_{r,\varphi,z}$ 
have a weak dependence on $n$, then the 
high-frequency ($n\gg1$) limits of 
equations (\ref{eq:15c}) and (\ref{eq:15d}) assume the simple forms
\begin{equation}
\vert{\cal P}_\parallel\vert^2\simeq\vert{\bar s}_r\vert^2\cos^2(\psi-\gamma_\parallel),
\label{eq:15f}
\end{equation}
and 
\begin{eqnarray}
\gamma_\parallel&\simeq&\left(\frac{2\Omega}{n\omega}\right)^{1/2}\frac{\Im\{{\bar s}_r{{\bar s}_\varphi}^*\}}{\vert{\bar s}_r\vert^2}\nonumber\\
&\simeq&[2\Omega/(n\omega)]^{1/2}[a_\parallel+b_\parallel(n-n_0)],
\label{eq:15g}
\end{eqnarray}
where 
\begin{equation}
a_\parallel\equiv\frac{\Im\{{\bar s}_r{{\bar s}_\varphi}^*\}}{\vert{\bar s}_r\vert^2}\Bigg\vert_{n=n_0},
\label{eq:15h}
\end{equation}
and
\begin{equation}
b_\parallel\equiv\frac{\partial}{\partial n}\left[\frac{\Im\{{\bar s}_r{{\bar s}_\varphi}^*\}}{\vert{\bar s}_r\vert^2}\right]\Bigg\vert_{n=n_0}.
\label{eq:15i}
\end{equation}
denote the first two coefficients in the 
Taylor expansion of 
$\Im\{{\bar s}_r{{\bar s}_\varphi}^*\}/\vert{\bar s}_r\vert^2$ 
about a local value $n_0$ of $n$.

\begin{figure}
\includegraphics[width=8cm]{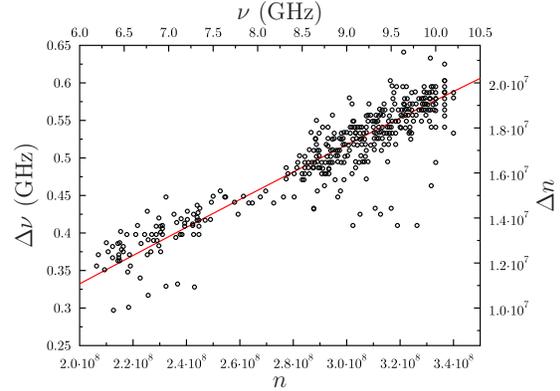}
\caption{The spacing $\Delta n$ (right axis) of the emission 
bands shown in Fig.~\ref{fig3} versus their mean 
harmonic number $n$ (lower axis).
The solid line represents the 
function given in equation (\ref{eq:15e}) for 
$\Omega/\omega=1.9\times10^4$, $a_\parallel=1.03\times10^4$ 
and $b_\parallel=1.48\times10^{-5}$.
The upper axis shows observation frequecy
$\nu = n\omega/2\pi$ relevant for the
Crab pulsar; the left axis shows $\Delta \nu$. 
Data points for the Crab pulsar
are taken from Fig.~10 of \citet{b14}.}
\label{fig4}
\end{figure}

In this case, equation (\ref{eq:15e}), together 
with equations (\ref{eq:14c}) and (\ref{eq:15f})--(\ref{eq:15i}), 
yields explicit expressions both for 
the radiated power $\vert{\cal P}_\parallel\vert^2$ and 
for the spacing $\Delta n$ between its successive peaks.
These expressions are plotted in Figs.~\ref{fig3} 
and \ref{fig4} over the 
interval $2\times 10^8\le n\le3.5\times 10^8$.  
In the case of the Crab pulsar, where the rotational 
frequency $\omega/(2\pi)$ is approximately $30$~Hz, 
the interval of harmonic numbers shown in Figs.~\ref{fig3} 
and \ref{fig4} 
corresponds to a frequency [$\nu=n\omega/(2\pi)$]
interval  covering $6$ to $10.5$~GHz,
i.e.\ to the interval of frequencies 
over which \citet{b14} observe the emission bands.
 
We have chosen a value $(1.9 \pm 0.1)\times10^4$ 
for the more fundamental free parameter 
$\Omega/\omega$ that renders the slope 
${\rm d}\Delta n/{\rm d}n$ predicted by 
equation (\ref{eq:15e}) the same as the slope
$0.058\pm0.01$ of the line passing through the 
observational data~\citep{b14} even when the phase 
$\gamma_\parallel$ is zero.\footnote{Note that the non-integral value of $\Omega/\omega$ can be approximated by an integer everywhere except in the factor $\sin(\pi\Omega/\omega)$ that appears in equation (\ref{eq:7}).} 
We have then fitted the $\Delta n$ intercept 
of the resulting line to that of the data, 
without changing its slope, by fixing the 
values $a_\parallel=(1.48\pm0.05)\times10^{-5}$ 
and $b_\parallel=(1.03\pm0.1)\times10^4$ 
of the parameters that appear 
in the expression for $\gamma_\parallel$.  

The Airy functions in equation (\ref{eq:14}) stop 
oscillating once the value of their 
argument falls below unity \citep{b19}.
We shall see in Sec.~\ref{newsec6} that the value 
$(1.9\pm0.1)\times10^4$ of the parameter 
$\Omega/\omega$ thus implied by the data 
of \citet{b14} places the last peaks of 
these functions at a frequency 
[$\sim(\Omega/\omega)^3\omega/\pi$] that agrees 
well with the position $\nu\sim 10^{15}$ Hz 
of the ultraviolet peak in the observed 
spectrum of the Crab pulsar \citep{b15}. 
The widths of the emission bands shown in 
Fig.~\ref{fig3} also agree with the widths of those 
that are observed:  the number of bands falling 
within the interval $2.7\times10^8\le n\le 3.5\times10^8$ 
in Fig.~\ref{fig3} is the same as the number of 
bands that occur within the corresponding 
frequency interval ($8.25$--$10.5$ GHz) in 
Figs.~6--8 of \citet{b14}.   

The contribution towards the radiated power 
from the component of the field perpendicular 
to the plane of rotation can be written, 
in analogy with equations (\ref{eq:15c}) and (\ref{eq:15d}), as 
\begin{eqnarray}
\vert{\cal P}_\perp\vert^2
&=&\sec(2\gamma_\perp)\big\{-\vert({\bar s}_\varphi\cos\theta_P-{\bar s}_z\sin\theta_P)\vert^2\sin^2\gamma_\perp\nonumber\\
&&+\frac{2\Omega}{n\omega}\vert{\bar s}_r\vert^2\cos^2\theta_P\cos^2\gamma_\perp+\big[\vert{\bar s}_\varphi-{\bar s}_z\sin\theta_P\vert^2\nonumber\\
&&-\frac{2\Omega}{n\omega}\vert{\bar s}_r\vert^2\cos^2\theta_P\big]\cos^2(\psi-\gamma_\perp)\big\},
\label{eq:16}
\end{eqnarray}
where
\begin{equation}
\gamma_\perp=\textstyle{1\over2}\arctan\frac{[2\Omega/(n\omega)]^{1/2}\Im\{({\bar s}_\varphi\cos\theta_P-{\bar s}_z\sin\theta_P){{\bar s}_r}^*\cos\theta_P\}}{\vert{\bar s}_\varphi\cos\theta_P-{\bar s}_z\sin\theta_P\vert^2-[2\Omega/(n\omega)]\vert{\bar s}_r\vert^2\cos^2\theta_P}.
\label{eq:17}
\end{equation}
These expressions imply a band spacing with essentially 
the same characteristics as those discussed above, 
differing from those in  equations (\ref{eq:15c}) 
and (\ref{eq:15d}) only in that ${\bar s}_r$ 
and ${\bar s}_\varphi$ exchange their roles 
with ${\bar s}_\varphi\cos\theta_P-{\bar s}_z\sin\theta_P$ 
and ${\bar s}_\varphi\cos\theta_P$, respectively.    

It will therefore be seen that the salient parameter
responsible for the frequency bands in the Crab pulsar
is the ratio $\Omega/\omega=(1.9\pm0.1)\times10^{4}$.
Using $\omega/2\pi \simeq 30$~Hz relevant for 
the Crab
yields a modulation frequency 
$\Omega/2\pi \simeq 570$~kHz.
A plausible cause for this modulation
is plasma oscillations within the
pulsar's atmosphere. A plasma frequency
of 570~kHz implies a free-electron
density $N_-\simeq 4\times 10^3~{\rm cm}^{-3}$,
a value that is consistent with the inferred
properties of the atmospheres 
of neutron stars.  

\section{Suppression of the emission bands by interference:
application to the main pulses of the Crab} 
\label{newsec5}
We now turn to the main pulses of the Crab pulsar,
in which the frequency-banded structure
is much less apparent~\citep{b14}. 
If the components of the electric current 
within the emitting region (the region in which 
the rotating distribution pattern of the 
current approaches the observer with the 
speed of light and zero acceleration) are 
such that ${\cal P}_\parallel$ and 
${\cal P}_\perp$ simultaneously contribute 
towards the value of the field, destructive 
interference of the contributions with 
differing polarizations could result 
in the suppression of the emission bands.  

This can be easily seen from equation (\ref{eq:15}) 
if we note that, in this case, the radiated power 
for $n\gg1$ is proportional to
\begin{equation}
\frac{{\rm d}P_n}{{\rm d}\Omega_P}\propto{Q_{\hat\varphi}}^2n^{-1/2}[\cos^2(\psi-\gamma_\parallel)+\kappa(n)\cos^2(\psi-\gamma_\perp)],
\label{eq:20}
\end{equation}
where 
\begin{equation}
\kappa(n)\equiv\Big\vert\frac{{\bar s}_\varphi\cos\theta_P-{\bar s}_z\sin\theta_P}{{\bar s}_r}\Big\vert^2
\label{eq:21}
\end{equation}
is a coefficient that, like $\gamma_\parallel$ 
and $\gamma_\perp$, could be expanded into a 
Taylor series about a local value $n_0$ of $n$.

\begin{figure}
\includegraphics[width=8cm]{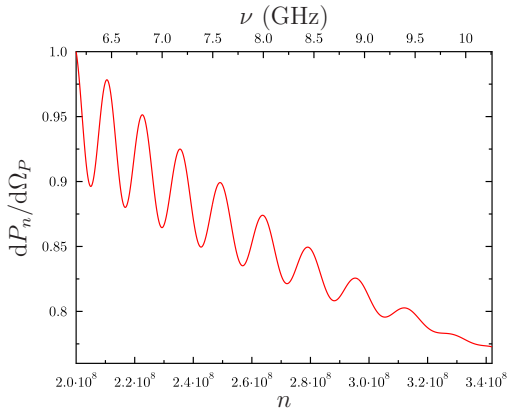}
\caption{The radiated power ${\rm d}P_n/{\rm d}\Omega_P$ per 
harmonic per unit solid angle (in arbitrary units) 
versus $n$ from a region of the magnetosphere in which the 
contributions from differing polarizations 
are comparable in magnitude.  
The dependence of the right-hand side of 
equation (\ref{eq:20}) on $n$ is shown for 
$Q_{\hat\varphi}\sim n^{-1}$, 
$\gamma_\perp=\gamma_\parallel+\pi/2$, 
$\kappa=0.9+0.15(n/n_0-1)$, and the same values of 
$\Omega/\omega$, $a_\parallel$, $b_\parallel$, 
and $n_0$ as in Fig.~\ref{fig4}.
The top axis of this Figure gives
frequency $\nu$ relevant for observations
of the Crab pulsar.}   
\label{fig5}
\end{figure}

Figure~\ref{fig5} shows the $n$ dependence of 
the right-hand side of equation (\ref{eq:20}) 
for a case in which ${Q_{\hat\varphi}}$ decays 
as $n^{-1}$, $\Omega/\omega=1.9\times10^4$ (as in Figs.~\ref{fig3} 
and \ref{fig4}), $a_\parallel=1.03\times10^4$, 
$b_\parallel=1.48\times10^{-5}$ (as in Fig.~\ref{fig4}), 
$\gamma_\perp=\gamma_\parallel+\pi/2$, 
and $\kappa$ can be approximated by $0.9+0.15(n/n_0-1)$ 
with $n_0=2\times10^8$.  
The emission bands are thus replaced by small-amplitude 
modulations of the radiation intensity that gradually die out.
The behaviour of the radiated power shown in Fig.~\ref{fig5} 
over the interval $2.7\times10^8\le n\le 3.5\times10^8$ 
is consistent with those of the 
short-lived microbursts that are observed over the 
corresponding frequency interval ($8.25$--$10.5$~GHz) 
within the main pulses of the Crab pulsar 
[see Figs.~2--4 of \citet{b14}].

The radiation described by equation (\ref{eq:14}) 
would be elliptically polarized if either 
${\bar s}_\varphi$ or ${\bar s}_r$ is dominant, 
and linearly polarized if ${\bar s}_z$ is dominant.  
The high degree of linear polarization of the 
interpulse in the frequency range 
$8.25-10.5$~GHz~\citep{b24} suggests, therefore, 
that the interpulse arises from a region of the 
magnetosphere in which the component of the 
electric current parallel to the rotation axis dominates, 
and so the radiation is polarized in the direction of 
${\hat{\bf e}}_\perp$, while the main pulse arises from 
a region in which more than one polarization 
contributes toward the observed field.
  
This explains not only the difference between 
the frequency structures of the main pulse and 
the interpulse, but also the difference between their 
temporal structures, i.e.\ the fact that the 
intensity of the interpulse has a broad and smooth 
distribution in time, while that of the main pulse 
is composed of randomly distributed nanoshots~\citep{b14}. 
It would be possible to identify the individual 
nondiffracting subbeams described in Sec.~\ref{newsec2} 
only in the case of a source whose length scale 
of spatial variations is comparable with 
${{\hat R}_P}^{-2}$ (in the case, e.g.\ of a 
turbulent plasma with a superluminally rotating 
macroscopic distribution).
The overall beam within which the 
nonspherically decaying radiation is detectable 
would then consist of an incoherent superposition of 
coherent, nondiffracting subbeams with widely 
differing amplitudes and phases, that would be 
observed as randomly distributed nanoshots.
Otherwise, the smooth transition between the 
adjacent filamentary parts of the source that 
generate neighbouring subbeams would result in an 
overall beam that is likewise smoothly distributed.
The filamentary sources sampled from an emitting 
region in which a single component of the electric 
current dominates would not be sufficiently different 
from one another in either amplitude or phase 
for their associated subbeams to be distinguishable.  

The observed difference between the dispersion measures 
of the main pulse and the interpulse \citep{b14} 
further supports the fact that these pulses arise 
from distinct regions of the magnetosphere.
As we have seen in Sec.~\ref{newsec2},  
an observer who is located at the polar angle 
$\theta_P$ samples, in the course of a rotation, 
the distribution of the density of the 
electric current around the cylinder 
${\hat r}=\csc\theta_P$.
Thus the profile of the mean pulse reflects
the azimuthal distribution pattern of the source.

\section{The broadband continuum spectrum of the radiation}
\label{newsec6}
\begin{figure}
\includegraphics[width=8cm]{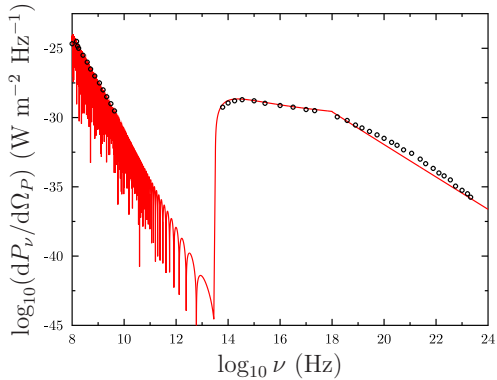}
\caption{The spectral distribution 
$\log({\rm d}P_\nu/{\rm d}\Omega_P)$, 
predicted by equation (\ref{eq:22}), 
versus $\log\nu$ for 
$\nu=n\omega/(2\pi)\simeq 30n$~Hz appropriate for the Crab pulsar. 
The points show observational data (where available)
of the spectrum of the Crab pulsar \citep{b15}.
In the model, the recovery
of intensity at the ultraviolet peak at $\sim 10^{15}$~Hz
is caused by resonant enhancement due to the azimuthal
modulation frequency $m\omega/2\pi \simeq 3$~THz.
The change in gradient at $\sim 10^{18}$~Hz
reflects a transition across the boundary of the 
Fresnel zone (see text).
}
\label{fig6}
\end{figure}

Since they only involve a limited range of frequencies, 
the features of the band emission we have discussed 
above do not depend on the spectral distributions of the 
Fourier transforms ${\bar s}_{r,\varphi,z}$ of the source 
densities $s_{r,\varphi,z}$ sensitively. 
To put the emission bands in the context of the 
continuum spectrum of the Crab pulsar, however, 
we would need both a more realistic model of 
the emitting current distribution and a more explicit 
description of the dependence of the density of 
emitting current on frequency.  
This notwithstanding, the frequencies across 
which the observed spectrum of the Crab pulsar 
undergoes sharp changes can be identified 
with the frequencies at which the Airy function 
and the coefficients $Q_{\hat\varphi}$ and 
${\bar Q}_r$ in the spectrum of the radiation 
described by equation (\ref{eq:14}) 
have their critical points. 
We have already seen that, for the value 
$\Omega/\omega = (1.9\pm0.1)\times10^4$ implied by 
the observational data of \citet{b14}, the 
position $\Omega^3/(\pi\omega^2)$ of the last peak of the 
Airy function in equation (\ref{eq:14}) 
coincides with the ultraviolet peak in the 
observed spectrum of the Crab pulsar \citep{b15}
(Fig.~\ref{fig6}).  

The coefficient $Q_{\hat\varphi}$ in equation (\ref{eq:14}) 
is independent of $n$ ($\sim2\Omega/\omega$) when $n\ll\mu_+$, 
decays as $n^{-1}$ when $n\gg\mu_+$, and is of the order 
of $n$ when $n\simeq m\gg\Omega/\omega$ 
[see equation(\ref{eq:7})].
Hence, if the frequency $m\omega/(2\pi)$ of the 
azimuthal variations of the source in the Crab pulsar 
[see equation (\ref{eq:1})] falls in the terahertz 
range, i.e.\ $m\sim10^{11}$, then the order of 
magnitude of $Q_{\hat\varphi}$ would change 
from $\Omega/\omega\sim10^4$ to $n\sim10^{11}$ across 
the terahertz gap in its observed spectrum. 
The corresponding enhancement of the radiated power by a 
factor of order $(\Delta{Q_{\hat\varphi}})^2\sim10^{14}$ 
is in fact consistent with observational data (see below
and Fig.~\ref{fig6}).
A plausible cause for a frequency $m\omega/2\pi \simeq 3$~THz
would be cyclotron resonance of free electrons in a
magnetic field of around $10^6$ G.
A magnetic field of this order at, or close to, the
light cylinder, is consistent with the
spin-down properties of the Crab pulsar~\citep{b15}.

Moreover, it can be seen from Equations (\ref{eq:8}) and 
(\ref{eq:8a}) that the coefficient ${\bar Q}_r$ 
changes from being independent of 
$n$ to decaying as $n^{-1/2}$ when 
$n$ increases past $\pi{\hat R}_P/({\hat r}_>-{\hat r}_<)^2$, 
and so the observation point falls within the Fresnel zone.
Given that the Crab pulsar is at the distance 
$R_P\simeq 7.1\times10^{21}$~cm and has a 
light cylinder with the radius 
$c/\omega \simeq 1.6\times10^8$~cm~\citep{b15}, 
this would account for the corresponding 
steepening of the observed spectrum of the 
Crab pulsar at $n\simeq3\times10^{16}$ 
(Fig.~\ref{fig6}) if the radial extent of the emitting plasma
observed from Earth 
has the value ${\hat r}_>-{\hat r}_<\simeq6.8\times10^{-2}$, 
i.e.\ $6.8$ percent of the light-cylinder radius.    

These generic features of the spectrum of 
the radiation from a superluminal source, 
which do not sensitively depend on the detailed 
distributions of the Fourier components 
${\bar s}_{r,\varphi,z}$ of the source densities,
are adequately described by the following simplified 
version of equation (\ref{eq:14}): 
\begin{equation}
\frac{{\rm d}P_n}{{\rm d}\Omega_P}\propto\vert{\bar Q}_r\vert^2{Q_{\hat\varphi}}^2{S(n)}^2\left(\frac{2}{n}\right)^{2/3}{\rm Ai}^2\left[-\left(\frac{2}{n}\right)^{1/3}\frac{\Omega}{\omega}\right],
\label{eq:22}
\end{equation}
where $S(n)$ stands for the dependence of 
the source density on $n$.
For $\Omega/\omega=1.9\times10^4$ and a 
${\bar Q}_r(n)$ that steepens by the factor 
$n^{-1/2}$ across $n\simeq3\times10^{16}$, 
equation (\ref{eq:22}) yields the continuum 
spectrum shown in Fig.~\ref{fig6} if we assume that 
$S(n)$ dominates over $Q_{\hat\varphi}$ everywhere, 
except across $n=10^{11}$ where 
$Q_{\hat\varphi}$ increases by the factor $10^{14}$
(due to the above-mentioned resonance with 
$m\omega/2\pi \simeq 3$~THz), 
and decays as $n^{-3/2}$ in $n<10^{11}$ 
and as $n^{-1/12}$ in $n>10^{11}$.
Note the close correspondence, in Fig.~\ref{fig6}, of 
equation (\ref{eq:22}) with the observational 
data on the spectrum of the Crab pulsar over 16 
orders of magnitude of frequency \citep{b15}.

\section{Discussion and Summary}
\label{newsec7}
We must stress the model-independent nature of
the results we have reported here.
The only global property of the magnetospheric 
structure of the Crab pulsar we have invoked 
is its quasi-steady time dependence: that the 
cylindrical components $j_{r,\varphi,z}(r,\varphi,z;t)$ 
of the density of the magnetospheric electric current 
depend on $\varphi$ only in the combination 
$\varphi-\omega t$.
Not only does this property follow from 
the observational data (as rigorously shown in 
Appendices A and B), but it is one that has been 
widely used in the published literature 
on pulsars [see, e.g., \citet{b23}]. 

Calculating the retarded field that is generated 
by the Fourier component associated with the 
frequency $\Omega$ of such a quasi-steady current 
distribution, we have derived a broadband radiation 
spectrum with oscillations whose peaks are proportionately 
spaced in, and decay algebraically with, frequency. 
By inferring the values of the parameters 
$\Omega/2\pi \simeq 570$~kHz 
and $m\omega/2\pi \simeq 3$~THz 
from the observational data and 
mildly restricting certain local properties 
of the source densities ${\bar s}_{r,\varphi,z}$ 
that appear in the derived expression 
we have {\em quantitatively} accounted for the 
following features in the observed spectrum of the Crab pulsar:
\begin{enumerate}
\item the spacing of the emission bands,
\item the frequency at which the extrapolated 
spacing between these bands would reduce to zero, 
\item the widths of the emission bands (their number 
within a given frequency interval),
\item the salient features of the continuum 
spectrum (its sharp rise across the terahertz gap, 
its ultraviolet peak, the change by $1$ of the value 
of its spectral index at $10^{18}$ Hz), and
\item the differences in the polarizations, 
dispersion measures, and the time and frequency 
signatures of the main pulse and the interpulse.
\end{enumerate}

Given that the derived band spacing $\Delta n$ 
increases as $n^{3/2}$ with the harmonic number $n$ 
[equations (\ref{eq:14c}) and (\ref{eq:15e})], 
we predict that the repetition of 
the observations of \citet{b14} at higher frequencies
would result in a dependence of the band spacing on 
frequency that has a local slope steeper than $0.058$.
There should be $12$ bands within the frequency 
interval $10$--$20$ GHz, for example, whose spacings 
would increase with frequency as
${\rm d}\Delta n/{\rm d}n\simeq0.086$.  

The two significant input parameters required
to simulate the Crab pulsar's spectrum are
a modulation frequency, $\Omega/2\pi \simeq 570$~kHz,
tentatively attributed to plasma oscillations of electrons
in the pulsar's atmosphere 
and $m\omega/2\pi \simeq 3$~THz, 
possibly corresponding to cyclotron resonance of electrons
in a magnetic field of around $10^6$~G.
The latter gives rise to the resonant
increase in spectral weight in the ultraviolet
(Fig.~\ref{fig6}).

At this point, it is worth
considering briefly the influence of
the ratio $\Omega/\omega$ on pulsar spectra
in general.
Given that typical features
of the pulsar's spectrum scale as
$(\Omega/\omega)^3$, we might 
expect that pulsars with slower
rotational frequencies (and therefore
small values of $\omega$) will have
observed spectral intensities weighted towards
higher frequencies. By contrast,
so-called millisecond pulsars 
(with large $\omega$) might
be expected to have emission
concentrated at lower frequencies than the Crab.
These predictions seem to be
borne out both by 
Geminga ($\omega/2\pi\simeq 4.5$~Hz),
which has emission peaked in the ultra-violet
to gamma-ray end of the spectrum,
and by millisecond pulsars such as
1937+21 ($\omega/2\pi \simeq 642$~Hz)
and 1957+20 ($\omega/2\pi \simeq 622$~Hz),
both of which show no emission in the GHz
range but strong pulses at 
radiofrequencies~\citep{b15}. 

Furthermore, once it is acknowledged 
that (as shown in Appendices A and B)
the current emitting the observed pulses from pulsars
has a superluminally-rotating distribution pattern, 
the results reported in the published literature 
on the electrodynamics of superluminal 
sources~\citep{b7,b8,b9,b10,b11,b12,b13} 
can be used to explain the extreme values of the giant pulses' 
brightness temperature ($\sim10^{40}$~K)~\citep{b10}, 
temporal width ($\sim 1$~ns) and source dimension 
($\sim 1$~m)~\citep{b12}, 
as well as the unique characteristics of the 
average pulses' polarization (their occurrence as 
concurrent `orthogonal' modes with swinging position 
angles and with nearly 100 per cent linear or 
circular polarization)~\citep{b11} and spectra 
(the range $-4$ to $-2/3$ of their spectral indices 
and the breadth of their bandwidth, 
from radio waves to gamma-rays)~\citep{b13}.

We conclude by re-emphasizing that
all these features of the observed radiation 
are consequences solely of the fact that 
the emitting current depends on
the coordinate $\varphi$
in the combination $(\varphi-\omega t)$. 
This constraint may be inferred directly from the
simplest observation, that pulsars
have a rigidly-rotating radiation distribution, 
and Maxwell's equations (see Appendices A and B);
it leads to a current distribution
with a superluminally rotating pattern
at a radius $r>c/\omega$ responsible
for the unique features of pulsar emission.
The explicit form of the current distribution
plays a role only in accounting for the pulse-to-pulse 
variations of the features 
discussed above (i.e.\ for 
the variable `weather' in the pulsar magnetosphere, 
not its stable `climate'). The plasma processes 
by means of which the magnetic field of the pulsar 
couples the rotational motion of the central 
neutron star to the observationally implied 
rigid rotation of the distribution pattern of 
the emitting current have no direct bearing on 
the results reported in this paper. 
The salient features of 
the observational data can be understood in terms 
of the superluminal emission mechanism 
without a knowledge of the magnetospheric structure of the pulsar.

\section{Acknowledgments}
We are grateful to Jean Eilek, Tim Hankins, 
Jim Sheckard, John Middleditch, Joe Fasel, 
and Andrea Schmidt for helpful discussions. 
This work is supported by U.S. Department of 
Energy Grant LDRD 20080085DR,
``Construction and use of superluminal emission 
technology demonstrators
with applications in radar, 
astrophysics and secure communications''.
A.A. also thanks the Royal Society for support.

\appendix

\section[]{Rigid rotation of the radiation pattern 
implies a rigid rotation of the source's distribution pattern}

As reflected in the highly stable periodicity of 
the mean profiles of the pulses detected on Earth, 
the pulsar radiation field has a rigidly rotating 
distribution pattern on average.
That is to say, the cylindrical components of the 
received radiation fields ${\bf E}$ and ${\bf B}$ 
depend on the azimuthal angle $\varphi$ only 
in the combination $\varphi-\omega t$:
\begin{equation}
E_{r,\varphi,z}(r,\varphi,z;t)=E_{r,\varphi,z}(r,\varphi-\omega t,z,t),
\label{eq:A1}
\end{equation}
\begin{equation}
B_{r,\varphi,z}(r,\varphi,z;t)=B_{r,\varphi,z}(r,\varphi-\omega t,z,t),
\label{eq:A2}
\end{equation}
where $(r,\varphi,z)$ are the cylindrical polar 
coordinates based on the axis of rotation, 
and $\omega$ is the angular frequency of 
rotation of the observed radiation pattern.
An equivalent statement is that the 
radiation fields ${\bf E}$ and ${\bf B}$ have a quasi-steady time dependence:
\begin{equation}
\left(\frac{\partial}{\partial t}+\omega\frac{\partial}{\partial\varphi}\right)E_{r,\varphi,z}=0,
\label{eq:A3}
\end{equation}
and
\begin{equation}
\left(\frac{\partial}{\partial t}+\omega\frac{\partial}{\partial\varphi}\right)B_{r,\varphi,z}=0;
\label{eq:A4}
\end{equation}
for the general solutions of these partial differential 
equations are given by the expressions 
on the right-hand sides of equations (\ref{eq:A1}) and (\ref{eq:A2}).

In the Lorenz gauge, the electromagnetic 
fields appearing in equation (\ref{eq:3}) are given 
by a four-potential $A^\mu$ that satisfies the wave equation
\begin{equation}
{\bf\nabla}^2A^\mu-{1\over c^2}{\partial^2A^\mu\over\partial t^2}=-{4\pi\over c}j^\mu,
\label{eq:A5}
\end{equation}
where $\mu=1,2,3$ designate the Cartesian
spatial components ${\bf A}$ and ${\bf j}$ of the potential 
$A^\mu$ and the current density 
$j^\mu$~\citep{b18}.  The retarded solution 
to this equation in unbounded space is 
given by equation (\ref{eq:4}), i.e.\ by 
\begin{equation}
A^\mu({\bf x}_P,t_P)=c^{-1}\int{\rm d}^3 x{\rm d}t\, j^\mu({\bf x},t)G({\bf x},t;{\bf x}_P,t_P),
\label{eq:A6}
\end{equation}
where
\begin{equation}
G({\bf x}, t;{\bf x}_P, t_P)={\delta(t_P-t-R/c)\over R}
\label{eq:A7}
\end{equation}
is the corresponding Green's function \citep{b18}.

Employing equation (\ref{eq:3}) to write 
out $B_{r,\varphi,z}$ in terms of the 
cylindrical components $A_{r,\varphi,z}$ of 
the vector potential, we see that the 
potential also has the time dependence 
expressed in equations (\ref{eq:A1})--(\ref{eq:A4}).  
It follows from $B_r=r^{-1}\partial A_z/\partial\varphi-\partial A_\varphi/\partial z$, for instance, that
\begin{equation} 
{\cal L}B_r={1\over r}{\partial\over\partial\varphi}{\cal L}A_z-{\partial\over\partial z}{\cal L}A_\varphi.
\label{eq:A8}
\end{equation}
where ${\cal L}$ stands for the differential operator appearing in equations (\ref{eq:A3}) and (\ref{eq:A4}):
\begin{equation}
{\cal L}\equiv{\partial\over\partial t}+\omega{\partial\over\partial\varphi}.
\label{eq:A8}
\end{equation}
This and the corresponding 
equations for $B_\varphi$ and $B_z$ show 
that equation (\ref{eq:A4}) is 
satisfied if ${\cal L}A_{r,\varphi,z}=0$, 
i.e.\ if $A_{r,\varphi,z}$ depend 
on $\varphi$ and $t$ only in the combination $\varphi-\omega t$.

On the other hand, the cylindrical components of the vector potential are related to the cylindrical components $j_{r,\varphi,z}$ of the current density via the following spatial part of equation (\ref{eq:A6}): 
\begin{eqnarray}
\lefteqn{\left[\matrix{A_{r_P}\cr A_{\varphi_P}\cr A_{z_P}\cr}\right] = c^{-1}\int r\,{\rm d}r\,{\rm d}\varphi\,{\rm d}z\,{\rm d}t\,G}\nonumber \\
                                                                 &   & \times\left[\matrix{\cos(\varphi_P-\varphi)j_r+\sin(\varphi_P-\varphi)j_\varphi\cr \cos(\varphi_P-\varphi)j_\varphi-\sin(\varphi_P-\varphi)j_r\cr j_z\cr}\right].
\label{eq:A9}
\end{eqnarray}
Applying the operator 
\begin{equation}
{\cal L}_P\equiv{\partial\over\partial t_P}+\omega{\partial\over\partial\varphi_P}
\label{eq:A10}
\end{equation}
to both sides of equation (\ref{eq:A9}) and making use of the 
fact that $G$ depends on $\varphi_P$ and $t_P$ 
in the combinations 
$\varphi_P-\varphi$ and $t_P-t$ [equation (\ref{eq:A7})],
and so ${\cal L}_PG=-{\cal L}G$, we find that 
the resulting equation can be cast in the form
\begin{eqnarray}
\lefteqn{{\cal L}_P\left[\matrix{A_{r_P}\cr A_{\varphi_P}\cr A_{z_P}\cr}\right]= c^{-1}\int r\,{\rm d}r\,{\rm d}\varphi\,{\rm d}z\,{\rm d}t\,G}\nonumber \\
                                                                          &  &\times\left[\matrix{\cos(\varphi_P-\varphi){\cal L}j_r+\sin(\varphi_P-\varphi){\cal L}j_\varphi\cr \cos(\varphi_P-\varphi){\cal L}j_\varphi-\sin(\varphi_P-\varphi){\cal L}j_r\cr {\cal L}j_z\cr}\right]
\label{eq:A11}
\end{eqnarray}
by means of integrations by parts with respect to $\varphi$ and $t$.  Hence, ${\cal L}A_{r_P,\varphi_P,z_P}$ vanishes if and only if ${\cal L}j_{r,\varphi,z}$ is zero.

It follows from equations (\ref{eq:A8}) 
and (\ref{eq:A11}), therefore, that a 
necessary and sufficient condition for 
the rigid rotation of the distribution 
pattern of the radiation field, 
i.e.\ ${\cal L}B_{r_P,\varphi_P,z_P}=0$, 
is the corresponding rigid rotation ${\cal L}j_{r,\varphi,z}=0$ 
of the distribution pattern of the source density.
 
\section{Rigid rotation of the source's 
distribution pattern extends beyond the light cylinder}

Our purpose here is to show that a solution 
of Maxwell's equations that has the quasi-steady time dependence
\begin{equation}
\frac{\partial}{\partial t}+\omega\frac{\partial}{\partial\varphi}=0,
\label{eq:B0}
\end{equation}
i.e.\ has a rotating distribution pattern, 
applies either to the entire magnetosphere 
or to an expanding region whose boundary 
propagates at the speed of light.  By considering 
the initial-boundary-value problem for the 
wave equation (\ref{eq:A5}) with Cauchy 
data satisfying equation (\ref{eq:B0}), 
we establish that such a solution cannot 
be smoothly matched to other types of solutions 
of Maxwell's equations across a boundary that 
is confined to a localized region of the magnetosphere.

The change of variables
\begin{equation}
\xi_1=r,\quad\xi_2=z,\quad\xi_3=\varphi-\omega t,\quad\xi_4=\psi(r,\varphi,z,t),\label{eq:B1}
\end{equation}
where $\psi$ is, for the moment, to be 
considered an arbitrary function, 
transforms the $z$-component of the wave equation (\ref{eq:A5}) into
\begin{eqnarray}
\lefteqn{\frac{1}{\xi_1}\left(\xi_1\frac{\partial A_z}{\partial\xi_1}\right)+\frac{\partial^2 A_z}{\partial{\xi_2}^2}+\left(\frac{1}{{\xi_1}^2}-\frac{\omega^2}{c^2}\right)\frac{\partial^2A_z}{\partial{\xi_3}^2}}\nonumber\\
&&+\left(\vert\nabla\psi\vert^2-\frac{1}{c^2}{\psi_t}^2\right)\frac{\partial^2A_z}{\partial{\xi_4}^2}+2\psi_r\frac{\partial^2A_z}{\partial\xi_1\partial\xi_4}\nonumber\\
&&+2\psi_z\frac{\partial^2A_z}{\partial\xi_2\partial\xi_4}+2\left(\frac{1}{r^2}\psi_\varphi+\frac{\omega}{c^2}\psi_t\right)\frac{\partial^2A_z}{\partial\xi_3\partial\xi_4}\nonumber\\
&&+\left(\nabla^2\psi-\frac{1}{c^2}\psi_{tt}\right)\frac{\partial A_z}{\partial\xi_4}=-\frac{4\pi}{c}j_z,
\label{eq:B2}
\end{eqnarray}
where $\psi_t\equiv\partial\psi/\partial t$, $\psi_r\equiv\partial\psi/\partial r$, etc.  The Jacobian of the above transformation is
\begin{equation}
\frac{\partial(\xi_1,\xi_2,\xi_3,\xi_4)}{\partial(r,z,\varphi,t)}=\frac{\partial\psi}{\partial t}+\omega\frac{\partial\psi}{\partial\varphi},
\label{eq:B3}
\end{equation}
which remains non-zero as long as $\psi$ is not a 
function that depends on $\varphi$ and $t$ as in $\varphi-\omega t$.

Now suppose that at a given time $t$ the 
surface $\psi(r,\varphi,z,t)=0$ represents 
the boundary of a region within which the 
solution to equation (\ref{eq:B2}) satisfies 
the symmetry (\ref{eq:B0}), i.e.\ within which 
$A_z$ is a function of $\xi_1$, $\xi_2$ and $\xi_3$ only.  
At all points on this boundary, we then have
\begin{equation}
\frac{\partial A_z}{\partial\xi_4}\bigg\vert_{\xi_4=0}=0
\label{eq:B4}
\end{equation}
and
\begin{eqnarray}
\lefteqn{\frac{1}{\xi_1}\frac{\partial}{\partial\xi_1}\left(\xi_1\frac{\partial A_z}{\partial\xi_1}\right)+\frac{\partial^2A_z}{\partial{\xi_2}^2}}\nonumber\\
&&+\left(\frac{1}{{\xi_1}^2}-\frac{\omega^2}{c^2}\right)\frac{\partial^2A_z}{\partial{\xi_3}^2}+\frac{4\pi}{c}j_z\bigg\vert_{\xi_4=0}=0,
\label{eq:B5}
\end{eqnarray}
where (\ref{eq:B5}) is the wave equation under symmetry (\ref{eq:B0}); (\ref{eq:B4}) and (\ref{eq:B5}) hold at the boundary of the region in question by virtue of holding inside that region.

Hence, any other type of solution of 
the wave equation that would match 
the symmetric solution across 
$\psi=0$ smoothly\footnote{A discontinuity in 
the value of $\partial A_z/\partial\xi_4$ that 
represents a field component is not permitted here, 
for it would entail the introduction of surface charges 
and currents with infinite densities within 
the magnetosphere.} has to be sought by solving the 
following initial-value problem: 
given that the Cauchy data on the hypersurface 
$\xi_4=0$ are those expressed in (\ref{eq:B4}) 
and (\ref{eq:B5}), what is the solution to 
the hyperbolic partial differential equation (\ref{eq:B2}) 
beyond $\xi_4=0$?  Note that since $\partial A_z/\partial\xi_4$ 
vanishes at {\em all} points of the hypersurface 
$\xi_4=0$, its derivatives $\partial^2A_z/\partial\xi_1\xi_4$, 
$\partial^2A_z/\partial\xi_2\xi_4$ and $\partial^2A_z/\partial\xi_3\xi_4$, 
which are in directions interior to this 
hypersurface, also vanish at $\xi_4=0$.  

Thus, equation (\ref{eq:B2}) in conjunction 
with the data (\ref{eq:B4}) and (\ref{eq:B5}) demands that 
\begin{equation}
\left(\vert\nabla\psi\vert^2-\frac{1}{c^2}{\psi_t}^2\right)\frac{\partial^2A_z}{\partial{\xi_4}^2}\bigg\vert_{\xi_4=0}=0.
\label{eq:B6}
\end{equation}
There are two ways in which this requirement could be met: 
either $\psi=0$ is a characteristic surface of the 
wave equation, and hence the first factor 
in (\ref{eq:B6}) vanishes, or at $\psi=0$ the 
derivative of $\partial A_z/\partial\xi_4$ in 
the direction {\em normal} to the boundary is also zero.
In the first case, the boundary of the domain in which 
symmetry (\ref{eq:B0}) is satisfied, $\psi=0$, will consist 
of an expanding wave front that propagates at speed $c$.
In the second case, the solution outside $\psi=0$ 
will also be symmetric, according to the 
Cauchy-Kowalewski theorem [cf.~\citet{b20}].
This is because the extension (by means of a Taylor series) 
of the data into an integral strip next to the 
boundary $\psi=0$ will yield a solution that is again 
independent of $\xi_4$.
Since the above argument applies also to the 
new boundary of the region thus extended, 
it follows that $A_z$ will also be be independent of 
$\xi_4$ outside the surface $\psi(r,\varphi,z,t)=0$, 
i.e.\ will be a function of $(r,\varphi-\omega t,z)$ 
throughout the magnetosphere.  

The corresponding results for $A_r$ and $A_\varphi$ 
may be derived in the same way.

\section{Asymptotic expansion of the spectrum 
for high radiation frequencies}

In this Appendix, we evaluate the leading term 
in the asymptotic expansion of the Anger 
functions that appear in equation (\ref{eq:6}) 
for a radiation frequency ($n\omega$) that appreciably 
exceeds both the rotation ($\omega$) and 
modulation ($\Omega$) frequencies of the source, 
though not necessarily the frequency 
$m\omega$ of its spatial oscillations.

In this regime, the Anger functions ${\bf J}_{n-\Omega/\omega}(n)$ 
and ${{\bf J}^\prime}_{n-\Omega/\omega}(n)$ 
can be approximated by the Bessel functions 
$J_{n-\Omega/\omega}(n)$ and  ${J^\prime}_{n-\Omega/\omega}(n)$, 
respectively [see equation (41) of \citet{b8}].
Once we cast these Bessel functions into the canonical 
forms $J_\mu(\mu+\zeta\mu^{1/3})$ and ${J^\prime}_\mu(\mu+\zeta\mu^{1/3})$ 
by introducing the new variables
\begin{equation}
\mu\equiv n-\frac{\Omega}{\omega},\qquad\qquad\zeta\equiv\frac{\Omega}{\omega}\left(n-\frac{\Omega}{\omega}\right)^{-1/3},
\label{eq:C1}
\end{equation}
we can write the leading terms in their asymptotic expansions for large $\mu$ as
\begin{equation}
J_\mu(\mu+\zeta\mu^{1/3})\simeq(2/\mu)^{1/3}{\rm Ai}(-2^{1/3}\zeta),
\label{eq:C2}
\end{equation}
and
\begin{equation}
{J^\prime}(\mu+\zeta\mu^{1/3})\simeq-(2/\mu)^{2/3}{\rm Ai}^\prime(-2^{1/3}\zeta)
\label{eq:C3}
\end{equation}
[see equations (9.3.23), (9.3.27) of \citet{b19}]. 
Taking the limits $n\gg\Omega/\omega$ of the 
coefficients and arguments of the above Airy functions, 
we obtain the expressions in equations (\ref{eq:12}) and (\ref{eq:13}).
   
For $(\Omega/\omega)\ll n\ll(\Omega/\omega)^3$, 
the argument of the resulting Airy function 
and its derivative that appear in equations (\ref{eq:12}) 
and (\ref{eq:13}) is large, so that we can further 
use the asymptotic approximations
\begin{equation}
{\rm Ai}(-\chi)\simeq\pi^{-1/2}\chi^{-1/4}\cos(\textstyle{2\over3}\chi^{3/2}-\pi/4),
\label{eq:C4}
\end{equation}
and
\begin{equation}
{\rm Ai}^\prime(-\chi)\simeq\pi^{-1/2}\chi^{1/4}\sin(\textstyle{2\over3}\chi^{3/2}-\pi/4),
\label{eq:C4a}
\end{equation}
[see equations (10.4.60) and (10.4.62) of \citet{b19}] 
to replace them by the trigonometric functions given 
in equations (\ref{eq:14a}) and (\ref{eq:14b}). 

Note that if $\Omega$ is replaced by $-\Omega$, 
the argument of the above Airy functions assume 
a positive value, and hence the resulting Anger 
functions ${\bf J}_{n+\Omega/\omega}(n)$ 
and ${{\bf J}^\prime}_{n+\Omega/\omega}(n)$ 
decay exponentially, rather than algebraically, 
with increasing $n$ [see \citet{b19}].

\end{document}